# Broadband ptychographic imaging for biological samples

**HUIXIANG LIN, AND FUCAI ZHANG***

*Department of Electrical and Electronic Engineering, Southern University of Science and Technology, Shenzhen 518055, China*

*zhangfc@sustech.edu.cn

**Abstract:** Ptychography is an attractive advance of coherent diffraction imaging (CDI), which can provide high lateral resolution and wide field of view. The theoretical resolution of ptychography is dose-limited, therefore making ptychography workable with a broadband source will be highly beneficial. However, broad spectra of light source conflict with the high coherence assumption in CDI that the current reconstruction algorithm were built upon. In this paper, we demonstrated that incorporation of a blind deconvolution in the reconstruction algorithm can improve the image quality of ptychography with broadband source. This broadband reconstruction algorithm can obtain high-quality amplitude and phase images of complex-valued samples requiring no knowledge of the illumination spectrum. Optical experiments using biological samples demonstrate the effectiveness of our method. The significant improvement in low coherence tolerance by our approach can pave the way for implementing ultrafast imaging with femtosecond or attosecond lasers or high-flux ptychographic imaging with laboratory EUV or X-ray sources.

## 1. Introduction

Ptychography is a scanning coherent diffraction imaging (CDI) technique that can image extended objects with high-resolution complex-value information [1-3]. In ptychography, a set of diffraction patterns is recorded by scanning with partially overlapping scan spots. The high degree of data redundancy leads to a fast convergence of reconstruction [1]. As a coherent imaging technique, ptychography is based on the assumption of full coherence of the source, which requires high spatial and temporal coherent sources. However, using filters to monochromate sources will result in a significant loss in photon flux, especially for the laboratory X-ray sources. This will affect the quality of ptychography whose resolution is dose-limited [4-5].

Recently, several numerical methods have been proposed to improve the tolerance of broad bandwidth which can obtain higher photon flux. Chen *et al.* proposed a multi-wavelength reconstruction algorithm in CDI experiments utilizing a tabletop high-harmonic-generation (HHG) source [6]. This method was also demonstrated by a synchrotron X-ray source, resulting in a 60-fold reduction in the exposure time compared with the monochromatic CDI [7]. The PolyCDI method shows a theoretical limit of 11% on the bandwidth. Moreover, this multi-wavelength method for X-ray ptychography has also been reported [8]. Another method also based on the coherent model is proposed [9] in CDI to extract monochromatic diffraction pattern from broadband diffraction data using a regularized inversion approach. The numerical monochromatization was also demonstrated in ptychography using femtosecond laser pulses [10]. A *priori* knowledge of the spectrum is necessary to perform these methods, which may increase the operational difficulty of experiments. Furthermore, another broadband reconstruction method introduces convolutional models into conventional CDI [11], which represents the diffraction pattern as a convolution of the coherent intensity and the Fourier transform of the complex coherence function. This method is demonstrated by an X-ray source in CDI [11] and achieves a satisfactory resolution at a bandwidth of 13.3% in CMI [12].

In this paper, we present an ultra-broadband ptychography technique that integrates a broadband convolution model, facilitating the imaging of dispersive samples without relying on prior spectral knowledge. The proposed approach is compatible with broadband sources across all wavelength ranges, including tabletop X-ray systems and high-harmonic generation (HHG) EUV sources.

## 2. Principle and Method

In ptychography, the object $O(\boldsymbol{r})$ is scanned by the probe $P(\boldsymbol{r})$ while the scan positions are represented by $\boldsymbol{s}_j$ ($j$ = 1,2,3,…, $J$, $J$ is the number of positions). At each position, the measured diffraction pattern in the far field with a monochromatic coherent illumination can be expressed

$$I_{mon}\left(\boldsymbol{q},\boldsymbol{s}_j\right)=\left|\mathcal{F}\left\{\varphi\left(\boldsymbol{r},\boldsymbol{s}_j\right)\right\}\right|^2, \tag{1}$$

where $\boldsymbol{r}$ is sample space coordinates and $\boldsymbol{q}$ is a reciprocal space coordinate. $\varphi(\boldsymbol{r},\boldsymbol{s}_j) = O(\boldsymbol{r},\boldsymbol{s}_j)P(\boldsymbol{r})$ is the complex exit wave of the sample. $\mathcal{F}$ represents the Fourier transform operator. High-contrast speckles of diffraction patterns are generated by the fully coherent illumination, while they will be blurred under broadband illumination. The effect of broad bandwidth can be included in the mutual coherence function (MCF), $\gamma$, which is equal to unity everywhere when the illumination is fully coherent. The broadband diffraction can be given as a convolution [13,14]

$$I_{broad}\left(\boldsymbol{q},\boldsymbol{s}_j\right)=I_{mon}\left(\boldsymbol{q},\boldsymbol{s}_j\right)\otimes\hat{\gamma}\left(\boldsymbol{q},\boldsymbol{s}_j\right), \tag{2}$$

where $\hat{\gamma}(\boldsymbol{q},\boldsymbol{s}_j)$ is the Fourier transform of the normalized MCF and $\otimes$ represents the convolution operator.

Our broadband ptychography introduces the convolution model and a Wiener-Lucy (W-L) tandem method we proposed in [12]. Similar to traditional ptychography, the estimated exit wave $\varphi(\boldsymbol{r},\boldsymbol{s}_j)$ from sample propagates to the detector plane, $\psi(\boldsymbol{q},\boldsymbol{s}_j)$, and then update its amplitude. However, before the modulus constraint, the estimated coherent diffraction intensity $I_{mon}(\boldsymbol{q},\boldsymbol{s}_j)$ is convolved with the $\hat{\gamma}(\boldsymbol{q},\boldsymbol{s}_j)$ to obtain a broadband estimate $I_{broad}(\boldsymbol{q},\boldsymbol{s}_j)$ as shown in Equation 2. Here, the $\hat{\gamma}(\boldsymbol{q}, \boldsymbol{s}_j)$ will be updated by performing W-L tandem deconvolution at every multiple of ten iterations,

$$\hat{\gamma}(\boldsymbol{q},\boldsymbol{s}_j) = deconvWL\left(I_{mon}(\boldsymbol{q},\boldsymbol{s}_j), I_M(\boldsymbol{q},\boldsymbol{s}_j)\right). \tag{3}$$

Then, the calculated wavefront is updated with measured intensity $I_M(\boldsymbol{q}, \boldsymbol{s}_j)$ by

$$\psi'(\boldsymbol{q},\boldsymbol{s}_j) = \psi(\boldsymbol{q},\boldsymbol{s}_j)\left[\frac{I_M(\boldsymbol{q},\boldsymbol{s}_j)}{I_{mon}(\boldsymbol{q},\boldsymbol{s}_j)}\right]^{1/2}. \tag{4}$$

The updated estimate $\psi'(\boldsymbol{q}, \boldsymbol{s}_j)$ is back-propagated to the object plane, giving the updated exit wave $\varphi'(\boldsymbol{r}, \boldsymbol{s}_j)$. Subsequently, apply the overlap constraint to update both the object and probe simultaneously according to the following functions,

$$O'(\boldsymbol{r},\boldsymbol{s}_j) = O(\boldsymbol{r},\boldsymbol{s}_j) + \alpha \frac{conj(P(\boldsymbol{r}))\left[\varphi'(\boldsymbol{r},\boldsymbol{s}_j) - \varphi(\boldsymbol{r},\boldsymbol{s}_j)\right]}{|P(\boldsymbol{r})|^2_{\max}}, \tag{5}$$

$$P'(\boldsymbol{r}) = P(\boldsymbol{r}) + \alpha \frac{conj\left(O(\boldsymbol{r},\boldsymbol{s}_j)\right)\left[\varphi'(\boldsymbol{r},\boldsymbol{s}_j) - \varphi(\boldsymbol{r},\boldsymbol{s}_j)\right]}{|O(\boldsymbol{r},\boldsymbol{s}_j)|^2_{\max}}, \tag{6}$$

where ‘conj’ denotes the complex conjugate operation. The coefficient $\alpha$, a constant set to a value of 1 in our experiment, could affect the convergence speed and robustness of the phase retrieval algorithm. Repeat the above steps until all positions have been updated to complete one iteration. The reconstruction process will terminate after a predetermined number of iterations.

## 3. Optical experiments

Our experimental setup is shown in Fig. 1, where a white LED was used as the light source. Figure 1(a) shows the spectra. After being collimated by an achromatic lens (Lens1), the broadband light passed through an aperture forming a probe. The light probe illuminated on the sample which is mounted on a stepper motor assemble and then the diffraction patterns of sample were recorded by a detector (Edge 4.2, PCO) with pixels 6.5 μm in size. Lens2 was used as a Fourier lens to achieve far-field propagation.

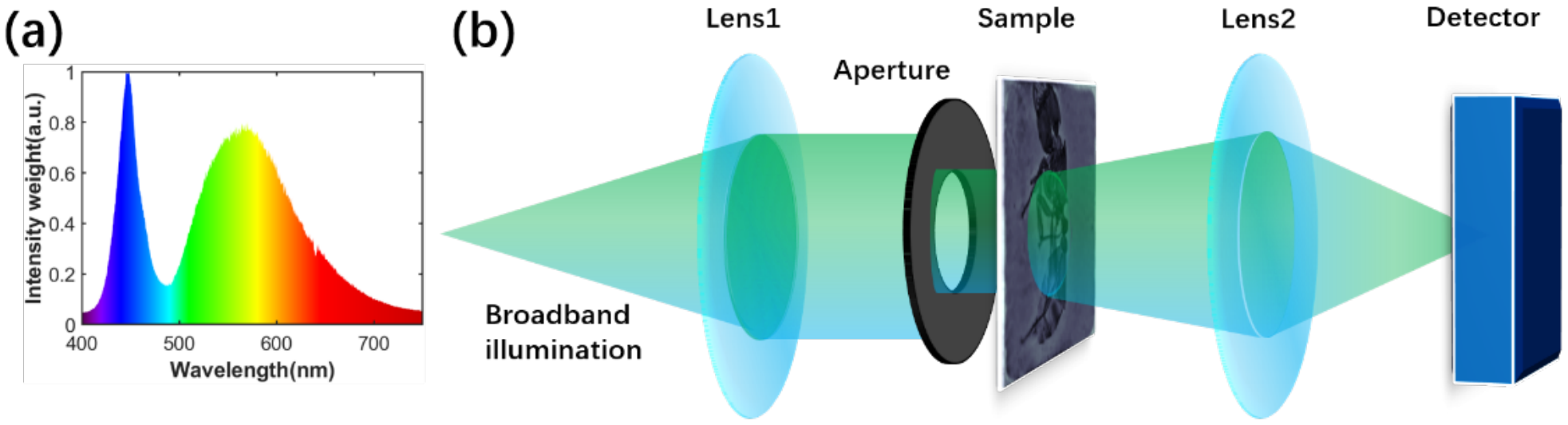

**Figure 1.** (a) The broadband spectra of a white LED. (b) Schematic of the light path.

A standard ant sample was used for imaging experimental verification. Here, the scanning of the sample followed a Fermat spiral trajectory which has been demonstrated that it can provide uniform coverage and high overlap ratio for a given number of scan points [15]. After data acquisition, traditional ptychography and our ultra-broadband ptychography algorithm were used to process the broadband diffraction data where the bandwidth of illumination is about 32%. Figure 2a illustrates the amplitude and phase of the ant sample reconstructed using traditional ptychography, where the details are no longer discernible. The reconstructed sample is overwhelmed by the improperly separated structured probe, leaving only a blurred outline. Our ultra-broadband ptychography, incorporating the convolution model, significantly enhances robustness to broadband illumination. Figure 2b shows the sample reconstructed using our method, where the amplitude exhibits sharp contours and reveals rich internal details. After phase unwrapping with the PUMA algorithm [16], the stacked structure of the ant is clearly resolved. This marks the first successful demonstration of clear ptychographic imaging of biological structures under an ultra-broadband source, validating the effectiveness of our method for imaging complex amplitude samples. This achievement significantly mitigates the limitations of broadband algorithms in practical biological and medical applications.

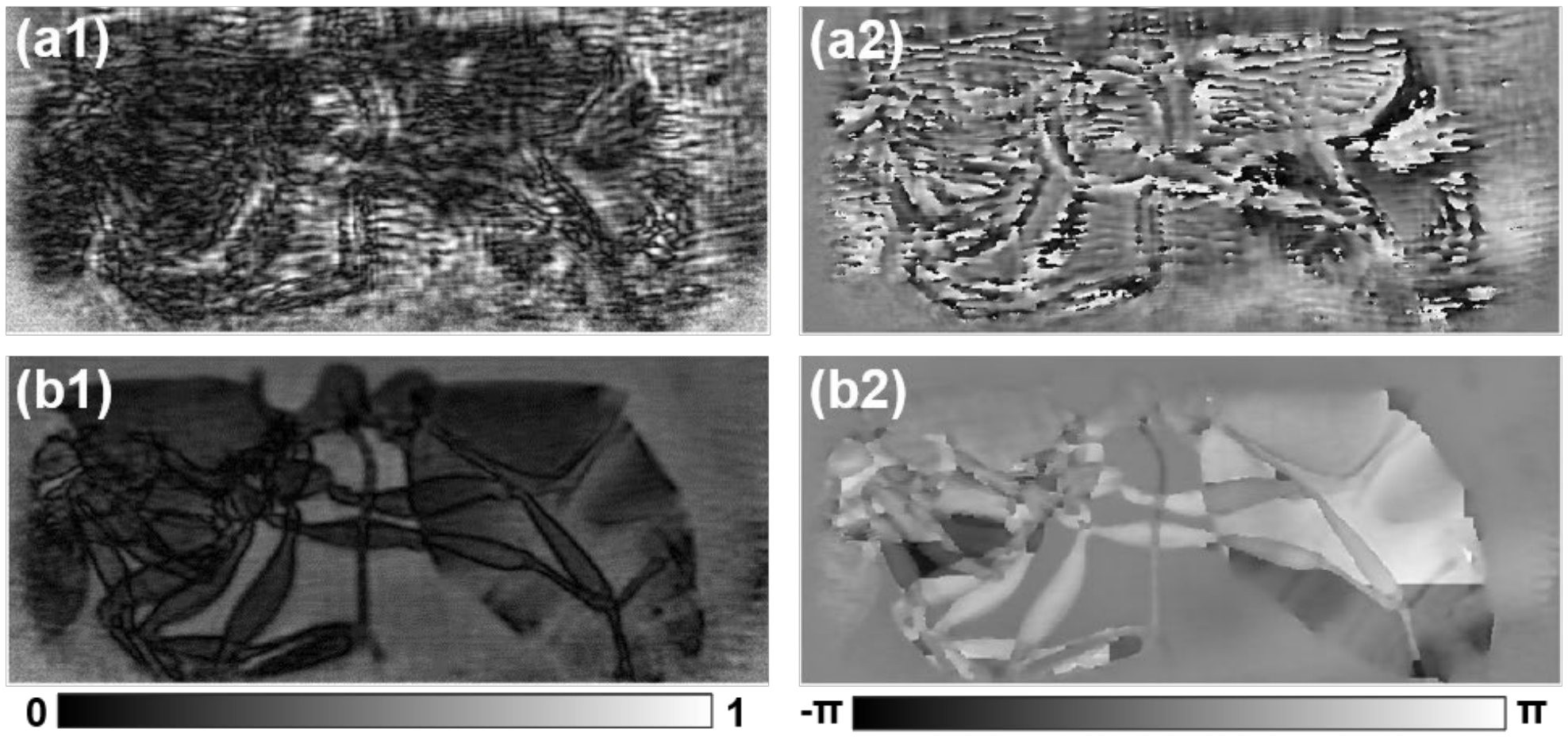

**Figure 2.** The amplitude (a1,b1) and phase (a2,b2) of an ant sample reconstructed by traditional ptychography and ultra-broadband ptychography respectively under white light source.

## 4. Conclusion

In summary, we present an ultra-broadband ptychography method capable of addressing dispersive samples, applicable to broadband sources across any wavelength range and samples with varying spectral responses. Our approach integrates the convolution model, enabling image reconstruction solely from broadband coherent diffraction patterns without requiring prior knowledge of the illumination spectrum. The effectiveness of our proposed method was verified by a visible light experiment using a source with 32% bandwidth. The phase structure information shown by the ant sample demonstrates our high-quality phase imaging capability. This will provide a new method for achieving high photon flux ptychography imaging. We believe that this broadband method will promote the applications of laboratory X-ray sources.